\documentclass[prl,showpacs,twocolumn]{revtex4}
\usepackage{array,amsmath,amssymb,graphicx}

\begin{document}



\title{Coherence Length of Cold Exciton Gases in Coupled Quantum Wells}

\author{Sen Yang, A.T. Hammack, M.M. Fogler, and L.V. Butov}
\affiliation{Department of Physics, University of California at San
Diego, La Jolla, CA 92093-0319}
\author{A.C. Gossard}
\affiliation{Materials Department, University of California at Santa
Barbara, Santa Barbara, California 93106-5050}

\date{\today}

\begin{abstract}
A Mach-Zehnder interferometer with spatial and spectral resolution
was used to probe spontaneous coherence in cold exciton gases, which
are implemented experimentally in the ring of indirect excitons in
coupled quantum wells. A strong enhancement of the exciton coherence
length is observed at temperatures below a few Kelvin. The increase
of the coherence length is correlated with the macroscopic spatial
ordering of excitons.

\end{abstract}

\pacs{78.67.-n,73.21.-b,71.35.-y}

\maketitle

Coherence of excitons in quantum wells attracts considerable
interest. It has been intensively studied by four-wave-mixing
\cite{Chemla}, coherent control \cite{Marie}, and interferometric
and speckle analysis of resonant Rayleigh scattering
\cite{Birkedal,Langbein,Haacke}. In all these experiments, exciton
coherence was induced by a resonant laser excitation and was lost
within a few ps after the excitation pulse due to exciton-exciton
and exciton-phonon collisions and due to inhomogeneous broadening by
disorder.

Another fundamentally interesting type of coherence is spontaneous
coherence (not driven by the laser excitation). Studies of
spontaneous coherence of excitons require implementation of cold
exciton gases, see below. This can be achieved with indirect
excitons in coupled quantum wells (CQW)~\cite{Butov04r}. Taking
advantage of their long lifetime and high cooling rate, one can
realize a gas of indirect excitons with temperature well below
$1\,\text{K}$ and density in excess of $10^{10}\,\text{cm}^{-2}$
~\cite{Butov04r}. For comparison, the crossover from classical to
quantum gas occurs at $T_{\rm dB}=2\pi \hbar^2 n/ (m g k_{\rm B})$
and $T_{\rm dB} \approx 3\,\text{K}$ for the exciton density per
spin state $n/g=10^{10}$cm$^{-2}$ (exciton mass $m = 0.22 m_0$, and
spin degeneracy $g=4$ for the GaAs/AlGaAs QWs \cite{Butov04r}). Note
that at this density $n a_B^2 \sim 0.1$ and, therefore, excitons are
interacting hydrogen-like Bose particles~\cite{Keldysh68} ($a_B
\approx 20$ nm is the exciton Bohr radius~\cite{Dignam}).

\begin{figure}[h] \includegraphics[width=8.5cm]{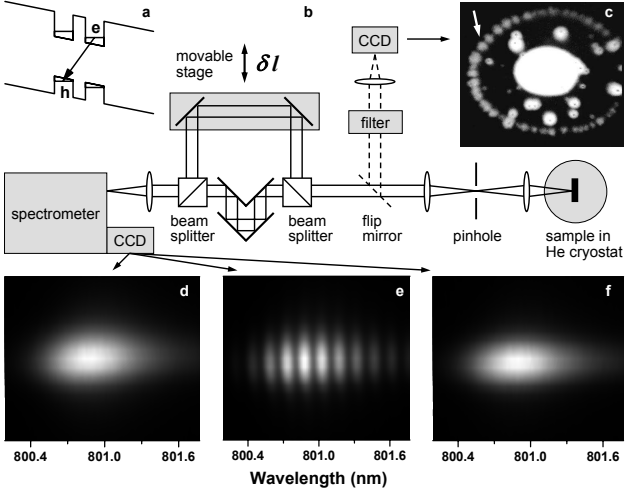}
\caption{
(a) The CQW band diagram. (b) Scheme of MZ interferometer with
spatial and spectral resolution.
(c) The pattern of indirect exciton PL. The area of view is $280 \times
250\, \mu$m. Spectra for the left (d), right (f), and both (e) arms of
the MZ interferometer. The light was selected from the center of the
arrow-marked MOES bead. The length of view (vertical axis) is $25\,
\mu$m. $T = 1.6$ K, $V_g = 1.24$ V, $D = 25\, \mu$m, $\delta l = 4.2$
mm, and $P_{ex} = 0.7$ mW for all the data.} \label{1}
\end{figure}

Spontaneous coherence can be experimentally studied using
nonresonant laser excitation so that coherence is not driven by the
laser. However, nonresonant excitation may lead to strong heating of
the excitons in the excitation spot \cite{Butov01}. Therefore, in
this paper we study coherence in the external exciton rings
\cite{Butov02}, which form far away from the excitation spot
(Fig.~\ref{1}c), at the border between the electron- and hole-rich
regions~\cite{Butov04,Rapaport}. The external ring of indirect
excitons in CQW is the region where the exciton gas is cold: The
excitons in the ring are formed from well-thermalized carriers and
their temperature essentially reaches that of the lattice. The cold
exciton gas in the external ring can form a macroscopically ordered
exciton state (MOES) --- an array of beads with spatial order on a
macroscopic length~\cite{Butov02}. The MOES appears abruptly along
the ring at $T$ below a few Kelvin.

Spontaneous coherence of excitons translates into coherence of the
emitted light~\cite{Ostereich,Laikhtman,Olaya-Castro,Zimmermann}. To
probe it several optical experimental techniques have been proposed:
a second-order optical response~\cite{Ostereich}, a
Hanbury-Brown-Twiss interferometry~\cite{Laikhtman,Olaya-Castro},
and a speckle analysis at off-resonant excitation~\cite{Zimmermann}.
Our technique is based on measuring the first-order coherence
function of the electric field $E(t, \textbf{r})$ of the light
emitted by excitons. This quantity is defined
by~\cite{Loudon,Scully}
\begin{equation}
g(t, \textbf{r}) = \langle E(t^\prime + t, \textbf{r}^\prime +
\textbf{r})
         E(t^\prime, \textbf{r}^\prime) \rangle /
        \langle E^2(t^\prime, \textbf{r}^\prime) \rangle.
\label{eq:g}
\end{equation}
(local ergodicity in space and time is assumed). The linear
technique allows us to work with the low level optical signals of
the spatially resolved photoluminescence (PL).

%
%
%

Our experimental setup (Fig.~\ref{1}b) is a variant of Mach-Zehnder
(MZ) interferometry with new ingredients. First, we added spatial
resolution by collecting the light only from a selected area of size
$D / M_1 = 2$--$10\,\mu\text{m}$ in the middle of a MOES bead (Fig.
1c). This was done by placing a pinhole of diameter $D =
10$--$50\,\mu\text{m}$ at the intermediate image plane of
magnification $M_1 = 5$. Second, we added the frequency resolution
by dispersing the output of the MZ interferometer with a grating
spectrometer. (The image was further magnified by the factor $M_2
\approx 2$ after the pinhole.) The output of the spectrometer was
imaged by a nitrogen cooled CCD (Fig.~\ref{1}b). The MZ delay length
$\delta l$ was controlled by a piezo-mechanical translation stage.
The PL pattern of the indirect excitons (Fig.~1c) was also imaged
with the pinhole removed and the image filtered at the indirect
exciton energy (dashed path in Fig.~\ref{1}b). The excitation was
supplied by HeNe laser at 633 nm (the laser excitation spot with
FWHM 7 $\mu$m is in the center of the exciton ring, Fig.~\ref{1}c).
The excitation was 400 meV above the indirect exciton energy and
well separated in space; therefore, no laser-driven coherence was
possible in the experiment. The CQW structure with two 8 nm GaAs QWs
separated by a 4 nm $\text{Al}_{0.33}\text{Ga}_{0.67}\text{As}$
barrier was grown by MBE, Fig.~\ref{1}a. For the applied external
gate voltage $V_g\approx1.2$ V the ground state is an indirect
exciton with a lifetime $\tau_{rec} \sim 40\,\text{ns}$ (details on
the CQW structures can be found in~\cite{Butov04r}).

%
%
\begin{figure}[h] \includegraphics[width=8.5cm]{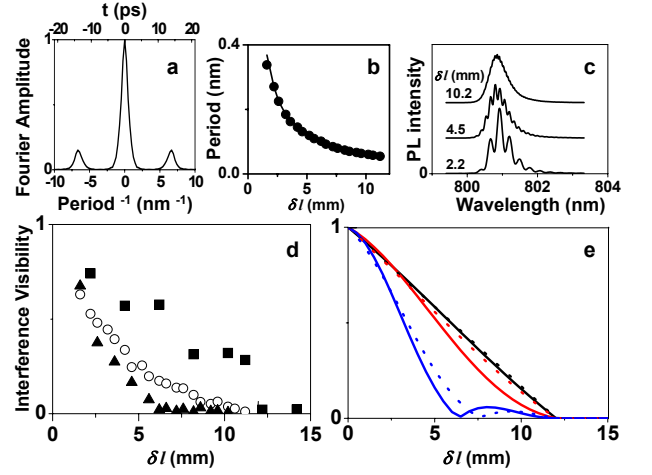}
\caption{
%
%
(a) The Fourier transforms of the CCD signal for $D = 50\,\mu$m and
$\delta l = 4.2$ mm. (b) Period of the interference fringes vs $\delta
l$. Solid line: fit to $\delta \lambda = \lambda^2/\delta l$. (c)
Interference profiles for $D = 50\,\mu$m and $\delta l = 2.2, 4.5,$ and
10.2 mm. (d) Measured and (e) calculated visibility of the interference
fringes vs $\delta l$ for $D = 50$ (triangles, blue), 25 (circles, red),
$10\,\mu$m (squares, black), and $M_2 = 1.7$. Solid and dotted lines in
(e) correspond to the Eqs.~(\ref{eq:V_general}) and
(\ref{eq:V_small_xi}), respectively. $T = 1.6$ K, $V_g =
1.24\,\text{V}$, $P_{ex} = 0.7$ mW for all the data.}
\label{2}
\end{figure}

An example of the measured interference pattern is shown in Fig.~1e.
The light was collected from the center of a bead shown in Fig.~1c
by the arrow. (While all interference profiles in the paper refer to
this spot, similar profiles were measured from other spots on the
ring.) The modulation period $\delta \lambda$ of the CCD signal $I$
was deduced from the locations of the satellite peaks of Fourier
transform of $I$, Fig.~2a. It was found to obey the expected
dependence $\delta \lambda = \lambda^2 / \delta l$ (see
Fig.~\ref{2}b and below). To quantify the amplitude of the
modulations we computed their visibility factor $V = (I_{\max} -
I_{\min}) / (I_{\max} + I_{\min})$ using a method based on the
Fourier analysis [Eq.~(\ref{eq:V})]. The visibility factor was
examined for a set of $\delta l$ and $T$.

%
%
\begin{figure}[h] \includegraphics[width=7cm]{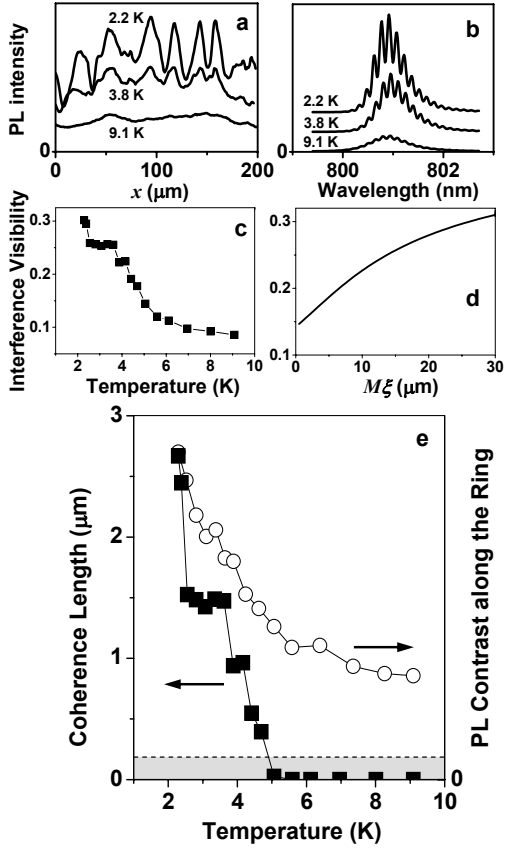}
\caption{
%
%
(a) Variations of the indirect exciton PL intensity along the
external ring at $T = 2.2, 3.8,$ and 9.1 K. (b) Interference
profiles at $T = 2.2, 3.8,$ and 9.1 K for $D = 50\,\mu$m and $\delta
l = 4.2$ mm. (c) Visibility of the interference fringes vs $T$. (d)
Calculated visibility as a function of the coherence length for $M_2
= 1.7$. (e) The exciton coherence length (squares) and contrast of
the spatial intensity modulation along the ring (circles) vs $T$.
The shaded area is beyond experimental accuracy. $V_g =
1.24\,\text{V}$, $P_{ex} = 0.7$ mW for all the data; $D = 50\,\mu$m,
and $\delta l = 4.2$ mm for the data in (b)-(e).} \label{3}
\end{figure}

The main experimental result is presented in Fig.~\ref{3}c:
Visibility of the interference fringes sharply increases at
temperatures below a few Kelvin. This contrasts with the
$T$-independent $V$ of the direct exciton emission measured at the
excitation spot center at $T = 2$--$10\,\text{K}$.

Let us proceed to the data analysis. Recall that for a classical
source with a Lorentzian emission lineshape, the first-order
coherence function [Eq.~(\ref{eq:g})] at the coincident points is
given by $g(t, \textbf{r} = 0) = \exp(-t / \tau_c)$, where $\tau_c$
is the inverse linewidth. By analogy, we assume the $r$-dependence
in the form
\begin{equation}
g(t, \textbf{r}) = g(t, 0) \exp(-r / \xi), \label{eq:xi}
\end{equation}
where $\xi$ is the coherence length. Our goal is to deduce $\xi$
from the contrast of the periodic modulations in the CCD image,
Fig.~1e. Consider the central row of that image. Let $x$ be a
coordinate along this row and let $x_0$ be the position of the
diffraction maximum for the central frequency of the emission line
$\omega_0 = 2 \pi c / \lambda_0$. Due to small width of this line,
it is permissible to work with small deviations $\delta x = x -
x_0$, $\delta \omega = \omega - \omega_0$, and $\delta \lambda =
\lambda - \lambda_0$. (Thus, the horizontal axes in Figs.~1d--f and
2c are labelled by the ``wavelength'' $\lambda$ using the conversion
formula $\delta \lambda /\lambda_0 = \delta x / x_0$.)

As mentioned above, the Fourier transform
\begin{equation}
\tilde{I}(t) = \int d x \exp\left(-i t \omega_0 x / x_0\right)
               I(x)
\label{eq:I_q}
\end{equation}
is found to possess three peaks: the main one, at $t = 0$, and two
satellites, Fig. 2a. We will show that these satellites occur at
$|t| = \tau = \delta l / c$. We will also explain the fact that the
shapes of the three peaks in Fig. 2a are nearly identical. Because
of the latter the amplitude of the oscillations in $I(x)$ is fully
characterized by the relative height of the main and the satellite
peaks. Therefore, we define the visibility factor by
\begin{equation}
            V = 2 |\tilde{I}(\tau)| / \tilde{I}(0).
\label{eq:V}
\end{equation}
Next, we note that the central row of the CCD image in Fig.~1d--f is
generated by the sources situated on the pinhole's diameter. Thus,
instead of two-dimensional vector $\textbf{r}$, it suffices to use
the linear coordinate $y$ along the magnified image of such a
diameter, of length $D_s = M_2 D$, at the spectrometer input slit.

The intensity of the CCD image averaged over a large time $T_{im}$
is a result of interaction of the original PL signal $E(t,y)$ with
two linear devices, the MZ interferometer and the spectrometer. It
is convenient to do the calculation of their combined effect in the
frequency domain. We define the Fourier amplitudes
$\tilde{E}(\omega_j, y) = \langle E(t, y) \exp(i \omega_j t)
\rangle$, for a set of frequencies $\omega_j = 2 \pi j / T_{im}$. A
straightforward derivation leads to
\begin{align}
I(x) &= \int\limits_{-D_s / 2}^{D_s / 2} \int\limits_{-D_s / 2}^{D_s
/ 2} d y_1 d y_2 \sum_{\omega_j}
 |1 + \exp(i \omega \tau)|^2
\notag\\
&\tilde{E}(\omega_j, y_1) \tilde{E}(-\omega_j, y_2) f_s(x, \omega_j,
y_1) f_s(x, \omega_j, y_2), \label{eq:I}
\end{align}
where
\begin{equation}
f_s(x, \omega, y) = \frac{\sin (\pi N z)}{\pi z}, \quad z =
\frac{\delta \omega - B y}{\omega_0} + \frac{\delta x}{x_0}
\label{eq:f_s}
\end{equation}
is the response function of the spectrometer, which is obtained from
the standard formula for the diffraction grating of $N$ grooves by
expansion in $\delta \omega$ and $\delta x$. Parameter $B$ is determined
by the linear dispersion of the spectrometer $A =
1.55\,\text{nm}/\text{mm}$, via the relation $B = 2 \pi c A /
\lambda_0^2$. After algebraic manipulations with Eqs.~(\ref{eq:g}),
(\ref{eq:I_q}), (\ref{eq:I}), and (\ref{eq:f_s}), we get the
following expression for the case of practical interest, $|t| < 2
\pi N / \omega_0$:
\begin{align}
\tilde{I}(t) & \propto \int\limits_{0}^{D_s} \frac{d y}{y t}
\sin\left[ \frac{1}{2} (D_s - |y|) B t\right]
\sin\left[\left(\frac{2 \pi N}{\omega_0} - |t|\right) \frac{B}{2}
y\right]
\notag\\
&\times \left[g(t, y) + \frac12 g(t - \tau, y) + \frac12 g(t + \tau,
y)\right].
\label{eq:tilde_I}
\end{align}
The three-peak structure of $\tilde{I}(t)$ described above stems
from the three terms on the last line of Eq.~(\ref{eq:tilde_I}). The
width of each peak is exactly the coherence time $\tau_c$. The peaks
are well separated at $\tau \gg \tau_c$ and their shape is nearly
identical if $\tau_c$ is sufficiently small. The heights
$\tilde{I}(0)$ and $\tilde{I}(\tau)$ of the peaks are determined by
the first and the second terms on the last line of
Eq.~(\ref{eq:tilde_I}), and so
\begin{gather}
V = \frac{(1 - \Delta) \int\limits_0^1 z^{-1} \sin[F (1 - \Delta) z]
\sin[F \Delta (1 - z)] \tilde{g}(z) d z} {{F \Delta} \int\limits_0^1
z^{-1} \sin(F z) \sin(1 - z) \tilde{g}(z) d z},
\notag\\
F \equiv \frac{\pi N A D_s}{\lambda_0},\quad
\tilde{g}(z) \equiv g\left(0, \frac{z D}{M_1}\right),\quad
\Delta \equiv \frac{\delta l}{  N \lambda_0}.
\label{eq:V_general}
\end{gather}
To understand the implications of this formula consider first the
case of an infinite diffraction grating, $N \to \infty$. Here
$\Delta \to 0$, $F \to \infty$ but the product $F \Delta
= \pi A D_s \delta l / \lambda_0^2$ remains finite. For the visibility
we get
\begin{equation}
V = {|\sin(F \Delta)|} / {F \Delta},
\label{eq:V_0}
\end{equation}
so that function $V(\delta l)$ has a periodic sequence of nodes at
$\delta l = n \lambda_0^2 / (A D_s)$, where $n = 1, 2,
\ldots$, and does not depend on $\xi$. Equation~(\ref{eq:V_0}) is
reminiscent of the Fraunhofer formula for diffraction through a slit
of width $D_s$.

In reality $N$ is large but finite. In this case the dependence on
$\xi$ does show up. Thus, for $M \xi \ll \lambda_0 / A N$,
where $M = M_1 M_2$, Eq.~(\ref{eq:V_general}) reduces to
\begin{equation}
V = \frac{1 - \Delta}{f \Delta} |\sin(f \Delta)|, \quad f =
\frac{\pi N A}{\lambda_0} (M_2 D - M \xi).
\label{eq:V_small_xi}
\end{equation}
To understand the origin of Eqs.~(\ref{eq:V_0}) and
(\ref{eq:V_small_xi}) consider the signal at the center of the CCD
image, at point $x_0$. It is created by interference between all
pairs of elementary input sources whose coordinates $y_1 = y +
\delta y$ and $y_2 = y - \delta y$ differ by no more than $\min\{M
\xi, \lambda_0 / A N\}$. What contributes to the image is the
Fourier harmonic of such sources shifted by $\delta\omega = B y$
with respect to the central frequency $\omega_0$,
cf.~Eq.~(\ref{eq:f_s}). The spread of $y$ across the pinhole results
into the spread of $|\delta\omega| \lesssim B (D_s - \delta y)$. If
$M \xi \ll \lambda_0 / A N$, then $D_s - \delta y = M_2 D - M \xi$
plays the role of the effective pinhole diameter in this
measurement. The resultant formula for visibility,
Eq.~(\ref{eq:V_small_xi}), is therefore similar to the Fraunhofer
formula for diffraction through a slit of this \emph{effective\/}
width.

We compared experimental $V(\delta l)$ with the above theory
treating $\xi$ and $M_2$ as adjustable parameters.
Instead of using the approximate Eq.~(\ref{eq:V_small_xi}), we
evaluated Eq.~(\ref{eq:V_general}) numerically. In agreement with
Eq.~(\ref{eq:V_small_xi}) $V$ was found to be most sensitive to
$\xi$ for $\Delta$ not too close to either zero or unity. It also
happened that $\xi$ was of the same order of magnitude as $\lambda_0
/ A N M$, and so the conditions for its estimation were nearly
optimal.

As seen in Fig.~2d,e, there is a good agreement between the theory and
the experiment. The measured $V(T)$, Fig. 3c, and the calculated
$V(\xi)$, Fig. 3d, allow us to obtain the coherence length $\xi(T)$.
Figure 3e shows that the coherence length increases sharply at $T$ below
a few Kelvin. Intriguingly, the increase of $\xi$ is in concert with the
MOES formation.

Naively, the interference pattern of an extended source of length
$\xi$ washes out when $\xi \delta k \sim \pi$, where $\delta k$ is a
spread of the momentum distribution. For $\xi \sim 2\, \mu$m (Fig.
3e), this gives $\delta k \sim 10^4$ cm$^{-1}$, which is much
smaller than the spread of the exciton momentum distribution in a
classical exciton gas $\delta k_{cl} \sim \hbar^{-1}\sqrt{2mk_BT}
\approx 3 \times 10^5$ cm$^{-1}$ at $T = 2$ K. In turn, this
corresponds to a spread of the exciton energy distribution $\hbar^2
\delta k^2/2m \sim 1\,\mu\text{eV}$, which is much smaller than that
for a classical exciton gas $\delta E_{cl} \sim k_B T \approx
200\,\mu\text{eV}$ at $T = 2$ K. It may also be interesting to
estimate the exciton phase-breaking time $\tau_\phi = \xi^2 / D_x$,
where $D_x \sim 10\,\text{cm}^2/\text{s}$~\cite{Ivanov_06} is the
exciton diffusion coefficient. Using again $\xi = 2\,\mu$m, we get
$\tau_\phi$ of a few ns. In comparison, the inverse linewidth
$\tau_c \approx 1\,\text{ps}$.

Let us now discuss physical mechanisms that may limit $\xi$ and
$\tau_\phi$. Since $\tau_{rec} \sim 40\,\text{ns} \gg \tau_\phi$,
the effect of exciton recombination on the phase-breaking time is
negligible. Next, the excitons are highly mobile, as evidenced by
their large diffusion length, ca. $30 \mu$m \cite{Levitov};
therefore, $\xi$ is not limited by disorder localization. The
coherence length may also be limited by inelastic collisions of
excitons with phonons and with each other. For the high exciton
densities $n \sim 10^{10}\,\text{cm}^{-2}$ in our experiments, the
dominant ones are the latter~\cite{Ivanov}. Note also that
spontaneous coherence we report arises in a cold thermalized exciton
gas (the lifetime $\tau_{rec}$ of the indirect excitons is much
longer than their thermalization time to $T = 2\,\text{K}$, $\sim 1$
ns \cite{Butov01}), and is therefore different from the laser-like
coherence in nonequilibrium systems due to a macroscopic coherent
optical field~\cite{Littlewood}.

Theoretical calculation of $\xi$ due to exciton interactions is yet
unavailable. It is expected however that inelastic processes should
vanish at $T = 0$. Our findings call for developing a quantitative
theory of phase-breaking processes in nonclassical exciton gases at
low temperatures when the thermal de Broglie wavelength is
comparable to the interparticle separation. In view of the exciting
phenomena uncovered in both fermionic~\cite{Altshuler} and
bosonic~\cite{Cornell:1995,Hulet:1995,Ketterle:1995} systems at low
temperatures, one can expect that rich physics may follow.

This work is supported by NSF grant DMR-0606543, ARO grant
W911NF-05-1-0527, and the Hellman Fund. We thank K.~L.~Campman for
growing the high quality samples, M.~Hanson, A.L.~Ivanov,
J.~Keeling, L.S.~Levitov, L.J.~Sham, B.D.~Simons, and A.V.~Sokolov
for discussions, G.O.~Andreev, A.V.~Mintsev, and E.~Shipton for help
in preparing the experiment.


\begin{thebibliography}{99}

\bibitem{Chemla}
D.S. Chemla and J. Shah, Nature {\bf 411}, 549 (2001).

\bibitem{Marie}
X. Marie, P.Le Jeune, T. Amand, M. Brousseau, J. Barrau, M.
Paillard, R. Planel, Phys. Rev. Lett. {\bf 79}, 3222 (1997).

\bibitem{Birkedal}
D. Birkedal and J. Shah, Phys. Rev. Lett. {\bf 81}, 2372 (1998).

\bibitem{Langbein}
W. Langbein, J.M. Hvam, and R. Zimmermann, Phys. Rev. Lett. {\bf
82}, 1040 (1999).

\bibitem{Haacke}
S. Haacke, S. Schaer, and B. Deveaud, V. Savona, Phys. Rev. B 61,
R5109 (2000).

\bibitem{Keldysh68}
L.V. Keldysh and A.N. Kozlov, Zh. Eksp. Teor. Fiz. {\bf 54} 978
(1968) [Sov. Phys. JETP {\bf 27}, 521 (1968)].

\bibitem{Butov04r}
L.V. Butov, J. Phys.: Condens. Matter {\bf 16}, R1577 (2004).

\bibitem{Dignam}
M.M. Dignam and J.E. Sipe, Phys. Rev. B 43, 4084 (1991).

\bibitem{Butov01}
L.V. Butov, A.L. Ivanov, A. Imamoglu, P.B. Littlewood, A.A.
Shashkin, V.T. Dolgopolov, K.L. Campman, and A.C. Gossard, Phys.
Rev. Lett. {\bf 86}, 5608 (2001).

\bibitem{Butov02}
L.V. Butov, A.C. Gossard, and D.S. Chemla, cond-mat/0204482; Nature
{\bf 418}, 751 (2002).

\bibitem{Butov04}
L.V. Butov, L.S. Levitov, A.V. Mintsev, B.D. Simons, A.C. Gossard,
D.S. Chemla, cond-mat/0308117; Phys. Rev. Lett. {\bf 92}, 117404
(2004).

\bibitem{Rapaport}
R. Rapaport, G. Chen, D. Snoke, S.H. Simon, L. Pfeiffer, K.West,
Y.Liu, S.Denev, cond-mat/0308150; Phys. Rev. Lett. {\bf 92}, 117405
(2004).

\bibitem{Ostereich}
Th. \"{O}stereich, T. Portengen, and L.J. Sham, Solid State Commun.
{\bf 100}, 325 (1996).

\bibitem{Laikhtman}
B. Laikhtman, Europhys. Lett. {\bf 43}, 53 (1998).

\bibitem{Olaya-Castro}
A. Olaya-Castro, F. J. Rodriguez, L. Quiroga, and C. Tejedor, Phys.
Rev. Lett. {\bf 87}, 246403 (2001).

\bibitem{Zimmermann}
R. Zimmermann, Solid State Commun. {\bf 134}, 43 (2005).

\bibitem{Loudon}
R. Loudon, {\it The Quantum Theory of Light}, 3d ed. (Oxford
University Press, 2000).

\bibitem{Scully}
M.O. Scully and M.S. Zubairy, {\it Quantum Optics} (Cambridge
University Press, 1997).

\bibitem{Ivanov_06} A.L. Ivanov, L.E. Smallwood, A.T. Hammack, Sen Yang,
L.V. Butov, and A.C. Gossard, Europhys. Lett. \textbf{73}, 920
(2006).

\bibitem{Levitov}
L.S. Levitov, B.D. Simons, L.V. Butov, Phys. Rev. Lett. {\bf 94},
176404 (2005).

\bibitem{Ivanov}
A.L. Ivanov, P.B. Littlewood, H. Haug, Phys. Rev. B {\bf 59}, 5032
(1999).

\bibitem{Littlewood} For a review, see
P.B. Littlewood, P.R. Eastham, J.M.J. Keeling, F.M. Marchetti, B.D.
Simons, and M.H. Szymanska, J. Phys.: Condens. Matter {\bf 16},
S3597 (2004).

\bibitem{Altshuler}
B.L. Altshuler, P.A. Lee, and R.A. Webb, {\it Mesoscopic Phenomena
in Solids}, edited by V.M. Agranonich and A.A. Maradudin
(North-Holland, Amsterdam 1991).

\bibitem{Cornell:1995}
M.~H. Anderson, J.~R. Ensher, M.~R. Matthews, C.~E. Wieman, E.~A.
Cornell, Science {\bf 269}, 198 (1995).

\bibitem{Hulet:1995}
C.~C. Bradley, C.~A. Sackett, J.~J. Tollett, R.~G. Hulet, Phys. Rev.
Lett. {\bf 75}, 1687 (1995).

\bibitem{Ketterle:1995}
K.~B. Davis, M.~O. Mewes, M.~R. Andrews, N.~J. Vandruten, D.~S.
Durfee, D.~M. Kurn, W.~Ketterle, Phys. Rev. Lett. {\bf 75}, 3969
(1995).

\end{thebibliography}
\end{document}